# Reassessment of the basis of cell size control based on analysis of cell-to-cell variability

Giuseppe Facchetti[1,*], Benjamin Knapp[2], Fred Chang[2,*], Martin Howard[1,*]

[1]John Innes Centre, Computational and Systems Biology, Norwich, United Kingdom

[2]University of California San Francisco, Cell and Tissue Biology, San Francisco, CA, USA

* Corresponding authors

| | |
|---|---|
| Giuseppe Facchetti | Giuseppe.Facchetti@jic.ac.uk |
| Fred Chang | Fred.Chang@ucsf.edu |
| Martin Howard | Martin.Howard@jic.ac.uk |

**ABSTRACT**

Fundamental mechanisms governing cell size control and homeostasis are still poorly understood. The relationship between sizes at division and birth in single cells is used as a metric to categorize the basis of size homeostasis [1-3]. Cells dividing at a fixed size regardless of birth size (sizer) are expected to show a division-birth slope of 0, whereas cells dividing after growing for a fixed size increment (adder) have an expected slope of +1 [4]. These two theoretical values are, however, rarely experimentally observed. For example, rod-shaped fission yeast *Schizosaccharomyces pombe* cells, which divide at a fixed surface area [5, 6], exhibit a division-birth slope for cell lengths of 0.25±0.02, significantly different from the expected sizer value of zero. Here we investigate possible reasons for this discrepancy by developing a mathematical model of sizer control including the relevant sources of variation. Our results support *pure* sizer control and show that deviation from zero slope is exaggerated by measurement of an inappropriate geometrical quantity (e.g., length instead of area), combined with cell-to-cell radius variability. The model predicts that mutants with greater errors in size sensing or septum positioning paradoxically appear to behave as better sizers. Furthermore, accounting for cell width variability, we show that pure sizer control can in some circumstances reproduce the apparent adder behaviour observed in *E. coli*. These findings demonstrate that analysis of geometric variation can lead to new insights into cell size control.

**SIGNIFICANCE**

How cells control their size is a fundamental but still open question. It has been assumed that the two principle methods of control, namely sizer (where cells always divide at a fixed size) and adder (where cells always add a fixed size increment), can be identified because the correlation between size at birth and size at division is equal to 0 and to +1, respectively. However, experiments mainly provide numbers in between these two values. We show that use of cell length as the size measure, together with cell-to-cell radius variability, can explain this discrepancy and support pure sizer control in fission yeast. Surprisingly, the same phenomenon might exclude adder control in bacteria like *E. coli*.

**INTRODUCTION**

Cell growth and cell division are coordinated during the cell cycle such that the size of a cell is maintained around a target value specific for each organism and cell type. However, how this outcome is achieved is still largely an open question. Investigations of cell size control intrinsically require a quantitative experimental approach, with many investigations focusing on the simple geometries of yeasts and bacteria. In fission yeast, analyses of individual cells show that they grow to a surface area of about 160 µm$^2$ before entering mitosis and dividing [6]. Phenomenological characterization of size homeostasis, measuring birth and division sizes of individual single cells, has shown that cells grow to a specified size regardless of birth size. These findings suggest that these cells monitor their own size, dividing at the same size regardless of the size at birth [1, 4, 7]. This principle is called sizer control. On the other hand, the same analysis on *E. coli* data reveals that these cells appear to add a constant size increment during each cell cycle [3, 8], so-called adder control. These cells show a positive correlation between size at birth and size at division [4, 9], so that shorter (longer) cells tend to divide shorter (longer). Theoretical studies have further investigated adder control in terms of robustness to stochastic perturbations and their consequences for the duration of different cell cycle phases [10, 11].

The interpretation of these measurements assumes an unambiguous correspondence between the observed behaviour (division-birth size correlation) and the underlying basis of size control. No correlation implies cells with pure sizer control; a perfect correlation implies cells with pure adder control. However, experimental data has revealed correlations that lie in between these two cases, results which have challenged the notion of a simple basis for size control. As a result, controversies over the basis of size control persist even in *E. coli* (sizer versus adder [12]), as well as in budding yeast (sizer versus adder [2, 13]), whereas a recent study has proposed a combination of a timer (fixed time duration cell cycle) and an adder for *C. crescentus* [14]. Because of its stereotypical shape, and greater available understanding, this work considers fission yeast as a reference model. Even in this case, the measured division-birth correlation is significantly different from zero, casting some doubt on the sizer hypothesis [15].

Previous work showed that size homeostasis in fission yeast is based on total surface area sensing (rather than on cell length or volume sensing) [5, 6]. Quantitative measurements support the idea that this surface area control is achieved by accumulation of Cdr2 in protein clusters ("nodes") in a cortical band around the nucleus. As a result of this localization, the Cdr2 local nodal density scales with cell surface area, which can then trigger mitosis via thresholding [5, 6]. Moreover, through use of a *cdr2-T166A rga2Δ* mutant, cell size homeostasis was successfully switched to length-based size control, confirming the key role of Cdr2 protein in the mechanism [6]. Critical to these conclusions were analyses of mutant cells with altered widths, using *rga2Δ* (thinner) and *rga4Δ* (fatter) mutants [6, 16, 17], which allowed for a robust distinction to be made between size controls based on length, area or volume. However, most data from the literature uses length as the measure of cell size [1, 7, 15] and for wild-type cells shows a significantly positive division-birth slope (approximately from 0.2 to 0.3), suggesting that cells might inherit and preserve some elements of size information from the previous cell cycle, similar to adder behaviour. Our data (Fig. 1A) shows a similarly positive division-birth slope of around 0.25. Clearly, these findings are not straightforwardly consistent with pure sizer control [6].

This non-zero slope raises the possibility that cells may display pure size control that is obscured by measurement errors or other sources of variability (Fig. 1B). Alternatively, cells might adopt an imprecise implementation or some "impure" mixture of adder and sizer controls (Fig. 1C). To test these possibilities, we sought to determine what division-birth correlations would be generated by pure sizer (or adder) control, if the relevant sources of variation were properly accounted for. Our first result is that we can explain all the correlations experimentally observed for fission yeast in terms of pure sizer control only. Experimentally measuring a geometrical measure of cell size (length) that is different to that intrinsically used by the cells in their size control (area), together with radius variability within the cell population, can generate a large shift of the division-birth slope to higher values. We surprisingly found that a reduction in size-sensing precision by the cell, or in the precision of placement of the division plane, can paradoxically appear to generate better sizer behaviour in a division-birth plot. These insights also apply to size control in *E. coli*, indicating that our approach may improve the assessment of cell size homeostasis control in many cell types.

## RESULTS
### Sources of variation and mathematical model

In order to improve our understanding of the variety of size homeostasis behaviours, we developed a simple mathematical model of pure area-based sizer control in fission yeast, including the relevant sources of variation in cell size regulation and measurement (Materials and Methods). Since a sizer mechanism is based on size sensing only, the model does not contain time as a variable and, therefore, we do not need to include any error in the duration of the cell cycle. We aimed to test whether such a model is sufficient to reproduce the apparently imperfect sizer behaviour observed above. We assumed that each source of variation is Gaussian distributed, with the free parameters being the following standard deviations: $\sigma_\mu$ (error $\mu$ that cells make in sensing size and therefore in mitotic commitment), $\sigma_\alpha$ (error giving asymmetric placement $\alpha$ of the division plane with respect to the exact mid-plane of the cell) and $\sigma_\rho$ (natural cell-to-cell variability $\rho$ of the radius). We also included $\sigma_\varepsilon$ as the standard deviation of the measurement error $\varepsilon$ (μm) due to image acquisition and segmentation (see Table 1 for the entire list of variables and parameters of the model). As an example, the measured cell radius ($R^*$, where in the following the mark * will denote a measured quantity) is given by $R^* = R(1 + \rho) + \varepsilon$, where $R$ is the mean radius of the cell population (see Materials and Methods). Similar expressions can also be derived for the length at birth and length at division. From these equations, we can calculate the coefficient of variation (CV) for all geometrical quantities at division (Materials and Methods). For instance, in case of the division length, the CV is given by the sum of three contributions:

$$\text{CV}^2_{L^*_d} \approx \sigma^2_\mu + \sigma^2_\rho + \left(\frac{2\pi R}{A}\right)^2 \sigma^2_\varepsilon. \tag{1}$$

Compared to simple probabilistic sizer models for which cells divide according to a probability of division as a function of cell size [2, 18], our equation describes the distribution of sizes at division as a combination of well-defined sources of error.

We first estimated the model parameters. All parameters, except for the noise $\sigma_\alpha$ due to asymmetry in division, were obtained by fitting to the measured CV of radius, length, area and volume at division (which are all clearly unaffected by $\sigma_\alpha$). We utilised images of wild-type cells acquired at different z-planes (to guarantee good focus). We then generated (and re-segmented) images at four different resolutions, i.e. with different values of the experimental error $\sigma_\varepsilon$. This was achieved by binning a square of *n* by *n* pixels into a single pixel, with *n* ranging from 1 to 4, i.e. four different values of $\sigma_\varepsilon$ (Fig. 2A). Cells were then segmented using an algorithm specific for rod or quasi-rod shaped cells (Materials and Methods). For each of the four geometrical quantities (radius, length, area and volume), and for each resolution, we measured the coefficient of variation of division size. We then fitted these data points with the model equations by tuning the 6 model parameters ($\sigma_\mu, \sigma_\rho$ and the four values of $\sigma_\varepsilon$). The minimal Root Mean-Square Deviation (RMSD) was obtained with $\sigma_\mu = 6.5\%$ and $\sigma_\rho = 2.7\%$ (Fig. 2B and S1). Finally, through measurement of the localization of the septum, we directly estimated the asymmetry in division in wild-type cells (Materials and Methods). By knowing $\sigma_\varepsilon$, we obtained the value of $\sigma_\alpha = 3.2\%$, consistent with literature reports [19] (Fig. 2C).

From the above estimation, we note the value of 6.5% for the area sensing error, which is smaller than the CV of the division length (about 7-8%) normally used to indicate the accuracy of size sensing [7]. This initial result indicates that the sizer mechanism in fission yeast is slightly more precise than previously thought.

**Alteration of cell radius variability supports pure sizer control**

Most of the analysis reported in the literature on rod-shaped cells is based on length, because length is the easiest quantity to measure. Therefore, it is important to find an appropriate way to interpret the outcomes from these measurements. In this case, our data for the wild-type alone shows a positive birth-division slope: 0.25±0.02 (Fig. 1A, S2, consistent with available values [1, 7, 15]). We next investigated whether these discrepancies were due to errors, and whether this data could support the existence of pure sizer control in fission yeast.

From the expressions of the measured length at birth ($L_b^*$) and measured length at division ($L_d^*$), while still having underlying area-based size control, we can derive an equation for the theoretical value of the division-birth slope (Materials and Methods):

$$\text{slope}(L_d^*, L_b^*) = \frac{2\sigma_\rho^2}{H^2 \sigma_\rho^2 + K^2}, \tag{2}$$

where $H^2 = 1 - \frac{4\pi R^2}{3A}$ and $K^2 = (\sigma_\mu^2 + \sigma_\alpha^2)/\left(1 - \frac{4\pi R^2}{3A}\right) + \left(\frac{4\pi R}{A}\right)^2 \sigma_\varepsilon^2 / \left(1 - \frac{4\pi R^2}{3A}\right)$. The expression defines a sigmoidal curve, with the radius variability having an important effect (Fig. S3A). The reasons for these dependencies are as follows. For surface-area based size control, both birth and division lengths strongly depend on the true value of the radius and on its variability from cell to cell, as cells with a bigger (smaller) radius divide shorter (longer) and give birth to shorter (longer) daughter cells, like in adder control. A consequence of this correlation is a positive division-birth slope (Fig. 3A), with a theoretical slope given by Eq. (2). This effect was already visible in Fig. 1A: due to area sensing, wider mutants have shorter birth and division lengths (large circles at the bottom-left of the plot), whereas thinner mutant have longer birth and division lengths (small circles at the top-right of the plot). On the contrary, when all the cells have the same radius (i.e. $\sigma_\rho = 0$), the true division length will be perturbed by the sensing error only, and will thus be independent from the length at birth, giving a zero division-birth slope.

To verify whether radius variation can explain the observed division-birth slope based on length, we exploited the natural variability of the radius that is present in wild-type cells. We modulated the variability of the radii of a population of wild-type cells by selecting subsets with a measured radius $R^*$ in the range $R \pm w$. For a given $w$, we extracted the division-birth slope and radius natural variability of the selected cells (Fig. S3B and Materials and Methods). By gradually reducing $w$ (i.e. by reducing the radius variability), the experimental division-birth slope decreased towards zero as predicted by the model, without any further parameter adjustment (Fig. 3B). We also analysed different combinations of wild-type, thin (*rga2Δ*) and fat (*rga4Δ*) cells. Each combination had a different variability of the radius and, according to our analysis, therefore also a different division-birth slope for length. Model predictions, again without any further parameter adjustment, were in good agreement with the experimental data (Fig. 3C). Overall, these results again support a *pure* area-based sizer control in fission yeast, where variations in the radius cause division-birth slopes based on length to deviate far from zero.

**Wild-type data alone supports area-based size control**

So far, our analysis has assumed surface area sensing. The previous experiments underpinning this result were based on variability of the cell radius introduced by thin (*rga2Δ*) and fat (*rga4Δ*) mutants [5, 6]. We attempted to overcome this limitation by again exploiting the natural variability of the radius in wild-type cells to verify that area sensing was preserved and was not an artefact due to use of the two mutants.

Following the same procedure used for area sensing, we developed a model based on a length sensing assumption and a model based on a volume sensing assumption (Materials and Methods). For each of the two models, we derived expressions for the CVs at division and the division-birth slopes for the various geometric quantities (length, area, volume). As for area sensing (Fig. 2B), the CV equations were used to estimate the parameters $\sigma_\varepsilon$, $\sigma_\mu$ and $\sigma_\rho$: by tuning their values (separately for length and volume), we found the best fit for the experimental division size CVs for wild-type cells (Fig. 3D). With the obtained parameter values (Table S1), the equations for the division-birth slopes were then used to calculate the theoretical value of the slopes. We found that the experimental CVs but not the experimental wild-type slopes can be reproduced by the model based on a volume sensing assumption, whereas the model based on a length sensing assumption was unable to reproduce either (Figs. 3D-E). On the other hand, the model based on area sensing provided a good fit to both.

Furthermore, for area-based size sensing, one would naively expect that the CV of area at division ($A_d^*$) should be smaller than that for length ($L_d^*$), i.e. $CV^2_{A_d^*} < CV^2_{L_d^*}$. Our data in Fig. 2B shows that this is only true for sufficiently small experimental measurement errors, $\sigma_\varepsilon$. This is again expected, since errors in measurement affect the area calculation more than for length, as radius only enters the area estimation. Only when $\sigma_\varepsilon$ is sufficiently small ($\sigma_\varepsilon < R\sigma_\rho$, as predicted by the model, see Materials and Methods) does area sensing emerge as being less variable.

These results indicate both the importance of precise measurements in order to correctly infer the basis of cell size control and, without reliance on mutants, that fission yeast sizer control is indeed based on surface area.

**Asymmetric division and defective size sensing can paradoxically appear to enhance sizer behaviour**
Besides experimental error and radius variability, the model also includes two further biological sources of variability: asymmetry and sensing precision. We next studied their role in size homeostasis behaviour, based on length measurements.

We showed above how variability in the radius can mask sizer behaviour, bringing the division-birth slope based on length far from zero (Fig. 3B-C). However, for a given value of the radius variability $\sigma_\rho$, Eq. 2 indicates that the slope decreases when the asymmetry ($\sigma_\alpha$) or the sensing ($\sigma_\mu$) error increases. This result means that increased errors in the division process (in term of septum positioning and/or size sensing) paradoxically generate better sizing behaviour in a division-birth plot, with a slope closer to zero (Fig. S3A).

These counter-intuitive behaviours can be rationalised as follows: in the case of a perfect sizer, an increase in the asymmetry error does not affect the distribution of the division length. As a consequence, the cloud of points of the division-birth size homeostasis plot is stretched along the x-direction only, causing a reduction of the slope (Fig. 4A). A bigger sensing error produces a wider distribution of both division and birth lengths, i.e. a stretch along both x- and y-directions. However, since the slope value is affected by (reduced) only by the horizontal stretch due to a wider distribution of birth lengths (Materials and Methods), this case turns out to be similar to the asymmetry case. In both cases, the wider spread of the data points reduces the statistical confidence of the slope value. Therefore, to attain similar levels of precision for the slope values as before, a higher number of segmented cells are required (Fig. S4A).

To verify these counter-intuitive predictions about the asymmetry ($\sigma_\alpha$) or the sensing ($\sigma_\mu$) error, we re-examined the case of the *pom1Δ* mutant cells, which divide shorter and often display asymmetric divisions as compared to wild-type cells ($CV_{L_b^*}$=22% vs 8.4% for the wild-type) [20, 21]. Although formally we cannot exclude that the *pom1* deletion might induce different geometric size sensing, we assume here that *pom1Δ* cells still use surface area for triggering mitosis. Moreover, because of the incompressibility of the G2 phase, cells that have a length at birth above 7-8 μm follow an adder/timer mechanism rather than a sizer mechanism [7] (grey colour in Fig. 4B). Therefore, to avoid interference from this altered size control and to exclude the effects of highly asymmetric divisions, we limited our analysis to cells that start the cell cycle at lengths below ~7 μm (green colour in Fig. 4B). For this subset, the CV of the division length is 9.0%, significantly higher than the wild-type (7.2%), indicating a higher sensing error by this mutant (estimated at $\sigma_\mu \approx 8.6\%$). In agreement with the value of 0.14 predicted by the model, the experimental division-birth slope decreases from +0.25 to +0.15±0.03 (Fig. 4B), making the mutant erroneously appear as a better sizer. For a repeat experiment on agar (Fig. S4B), with a division length CV of 12.2%, the model predicts a slope of 0.08 in good agreement with the experimental measurement of 0.09±0.07.

**Model predictions match length- and volume-based sizer control experiments**
In a previous work we manipulated fission yeast size sensing in order to generate Cdr2-dependent cell size sensing according to cell length, using the *cdr2-T166A rga2Δ* background. Following this reprogramming, cell division indeed occurred based on length rather than surface area [6]. According to our model for length-based size control, the division-birth slope for length should be zero regardless of the natural variability of the radius and of the segmentation error (Materials and Methods). The experimental value for *cdr2-T166A rga2Δ* of +0.03±0.05 confirms our model prediction and provides additional support for length sensing in this mutant (Fig. 4C).

In the same work, we showed that division in the *cdr2Δ* mutant is no longer controlled by area and is instead based on a secondary Cdr2-independent mechanism that moves towards volume control [6]. Similar to *pom1Δ*, data from this mutant can be split into two regimes: a sizer-like regime at shorter cell birth lengths (less than 10 μm) and an adder/timer-like regime for longer cell birth lengths. The division-birth slope obtained for the former regime is 0.47 ±0.04 (Fig. 4D), which, although still (just) closer to sizer than adder behaviour, is even further away from the ideal sizer behaviour than the wild-type. Although the CV of the division length (7.0%) is similar to the wild-type value, an important question to answer is whether *cdr2Δ* is still implementing a sizer.

For a model based on pure volume sensing, we can again derive the equation for the division-birth slope (Materials and Methods). The equation and the result indicate that the effect of natural variability of the radius is bigger for volume sensing than for area sensing (Fig. 3F). Intuitively, this result follows because cell volume has a quadratic dependence on the radius, whereas area has a linear dependence only. For this reason, the shift of the division-birth slope away from zero will be more pronounced. Without any parameter changing, the model prediction is a slope of +0.62, close to the observed value of around +0.5. Therefore, for *cdr2Δ*, as for all the cases analysed in this study (Fig. 4E), the appearance of more adder-like behaviour is likely due to the use of an incorrect geometrical feature (length) and to the effect of variability in the cell radius.

**Incorporating a varying radius into the mathematical model**

So far, the model was based on the assumption that cell radius does not change from birth to division. Experimental data, however, show a strong, but not perfect, positive correlation between the two measurements ($q = 0.72$, Fig. S5A), indicating that some adjustments in the cell width might occur during the cell cycle. As an additional test, we therefore relaxed the constant radius assumption and developed a generalised mathematical model varying radius from $R_b = R(1 + \rho)$ at birth to $R_d = q \cdot R_b + R(1 - q)$ at division. These dynamics guarantee the preservation through generations of cell radius around the average value $R$. Numerical simulations show that the key features of the model outcomes are preserved. Moreover, since the variation of radius within one cell cycle is relatively small (Fig. S5A), we did not find significant changes in our results. As a result, the agreement with the experimental evidence is still good (Fig. S5B).

**Adder behaviour can emerge from pure sizer control**

With the exception of length sensing in the *cdr2-T166A rga2Δ* mutant, utilising length as the experimental size measure partially masks pure sizer behaviour by moving the division-birth slope towards adder behaviour, i.e. towards a value of +1. We therefore asked whether the same effect is present in a pure adder mechanism. In particular, we focus on an adder mechanism based on cell volume. This assumption allowed us to develop a model (Materials and Methods) for the widely studied bacterium *E. coli* which shares the same rod-shape as fission yeast and which is believed to utilize cell volume as the key geometrical quantity for size control [22]. In contrast with other models [23], also in this adder model we did not include any direct errors on the time durations: rather, variation in the duration of the cell cycle is a consequence of the error in sensing the added size, which is the fundamental quantity monitored by a cell which implements adder control. Similar to our observations for pure sizer control based on volume sensing, the effect of natural cell radius variability causes a monotonic increase of the division-birth length slope from the theoretical adder value of +1 up to +2 (Fig. S6A). By using 3.5% as an estimation of the radius variability [9], our adder model generates a division-birth length slope of +1.51±0.03 (Fig. 5A). Robustness of the outcome was tested by additively increasing/decreasing each standard deviation by up to 1% (or 2%): slope values remained entirely above +1 (Fig. S6B). However, all the experimentally reported division-birth length slopes for *E. coli* range from +0.7 to +1 [8, 9], i.e. they never overshoot the theoretical pure adder value. This result may suggest that an adder mechanism is incompatible with the observed behaviour.

Our results on the *cdr2Δ* mutant revealed that, in the case of a volume-based pure sizer, the effect of radius variability is more pronounced than in the case of area sensing and moves the division-birth length slope much closer to adder-like behaviour. Running a numerical simulation of such a volume-based sizer model with $\sigma_\rho = 3.5\%$ (slightly higher than the radius variability in fission yeast) on *E. coli* sized cells [9], we obtained a division-birth length slope of +0.81±0.02 (Fig. 5B), in agreement with the experimental results from this bacterium. This finding strongly suggests that sizer behaviour in *E. coli* cannot currently be ruled out.

**DISCUSSION**

Phenomenological characterization of size homeostasis is a common approach for extracting information about cell size regulation. Although initial results previously suggested an imperfect or possibly mixed sizer-like control in fission yeast, we have shown here that *pure* sizer control based on *accurate* sensing of cell size is implemented in this organism, using a Cdr2-dependent mechanism with area-based control in the wild-type and length-based control in *cdr2-T166A rga2Δ* [6]. Moreover, we have identified three generalizable aspects in the analyses of size homeostasis: (i) high accuracy in the measurement of the cell size is necessary to avoid misleading data (Fig. 2B); (ii) use of the wrong geometrical quantity (e.g., length instead of area), combined with natural variability of the cell radius, distorts the observed size control behaviours (Fig. 3B-C). While this effect disappears in the presence of length-based sizer sensing (see *cdr2-T166A rga2Δ*, Fig. 4C), the behaviour is even more markedly perturbed when sizer control is based on cell volume (see *cdr2Δ* in Fig. 4D and *E. coli* in Fig. 5B); (iii) asymmetric cell division and, more surprisingly, a reduction in the precision of size sensing, can paradoxically appear to enhance sizer behaviour. As shown by the case of *pom1Δ* in fission yeast, an apparently stronger sizer behaviour (lower division-birth slope) may be due to a loss of precision in size control (Fig. 4B). These demonstrations show that correct interpretation of size homeostasis behaviour requires accurate measurements of multiple geometrical quantities. Furthermore, "intermediate" behaviour (as in wild-type cells) may not necessarily be due to imperfect size control, nor to a combination of two or more distinct size controls (e.g. sizer and adder). Even limited variability can move the division-birth length slope from 0 towards +1 in the case of pure sizer control and from +1 towards +2 in case of pure adder control, thus completely obscuring the underlying fundamental basis of size control (Fig. 6).

An important aspect that has emerged in our analysis is the effect of radius variability. One might expect that the small variations found from cell to cell might have only a marginal effect on size control behaviour. Surprisingly, the curve and data in Fig. 3C show that, in the case of area-based sizer sensing in fission yeast, a radius variability of only 7% is sufficient to move the division-birth slope close to +1, making pure sizer control wrongly appear as a pure adder. The equations of the model allow us to understand how any particular size homeostasis behaviour can emerge from antagonism between inaccuracy in the division process (asymmetry and size sensing) and radius variability. In the approximation of the slope expression given in Eq. (2), the first non-zero term is $2\sigma_\rho^2/K^2$. If we ignore the contribution of the experimental error $\sigma_\varepsilon^2$, this term can be approximated by $2\sigma_\rho^2/(\sigma_\alpha^2 + \sigma_\mu^2)$, which clearly displays the antagonism between the two effects. In our case, the combination of the estimated values for asymmetry ($\sigma_\alpha = 3.2\%$) and size sensing ($\sigma_\mu = 6.5\%$) errors gives a total standard deviation of 7.3%. Therefore, we can expect that a radius variability of the same order of magnitude can neutralize the effect of these two sources of variation and lead to adder-like behaviour.

For the same reason, despite the adder phenotype, sizer control may still not be ruled out in *E. coli* (Fig. 5), consistent with some alternative theories [12, 22, 24]. Furthermore, according to our analysis, adder control would tend to show behaviour closer to a timer (slope significantly above +1). One caveat is that our adder model considers the simple case of a constant cell size increment from birth to division. It has been alternatively suggested that a constant volume (per origin) is added between two initiations and that cells then divide a constant time after initiation [22]. Although further investigation is needed to better understand this misalignment between the basis of size control and the resulting division-birth slope behaviour, our results suggests that *E. coli* might not implement adder control.

Overall, the analysis presented here shows that pure sizer control is able to reproduce a large variety of cell size homoeostasis behaviours. Our results, with careful attention to both sources of variation in size control and to geometrical aspects preserved by the cell, could be widely relevant in deciphering size homeostasis behaviour in many cell types. Such accurate analyses of cell size behaviours will be critical for dissecting the underlying molecular mechanisms responsible.

**FIGURES AND TABLES**

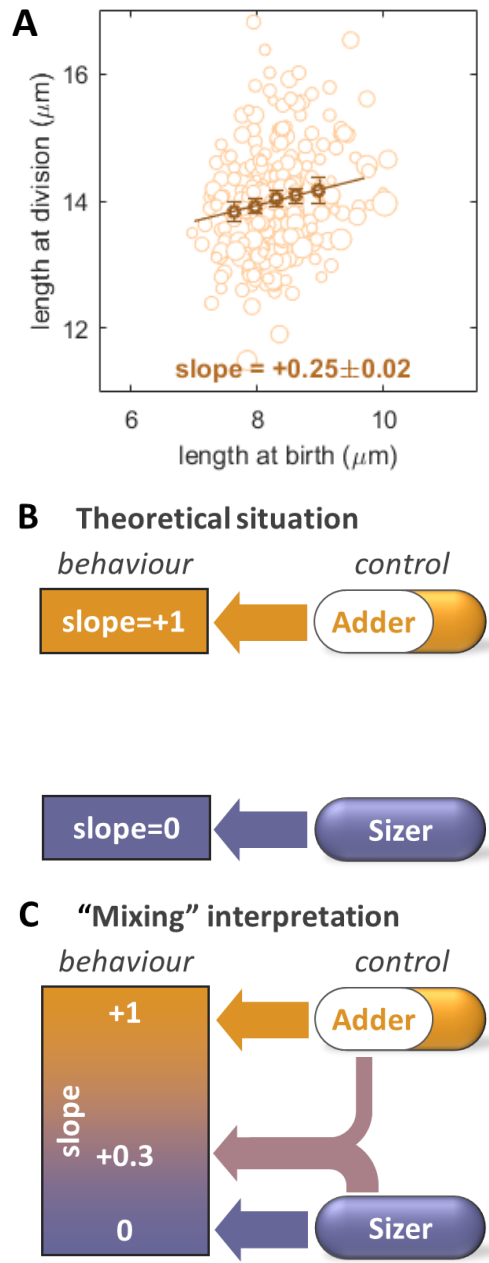

**Figure 1: Mapping between cell behaviour and underlying basis of size control.** (**A**) Division-birth correlation for wild-type (FC15, n=326) based on measurement of cell length. Size of circles proportional to cell radius. Binned data, with mean value ± standard error, also shown (dark circles), together with best fit line, with indicated slope. (**B**) Theoretical assumption of a one-to-one mapping between division-birth slope and size control, where slope=0 implies sizer control and slope=+1 implies adder control. No intermediate values for the slope are expected. (**C**) Slope values between 0 and +1 could be interpreted as a mix between sizer and adder control (or an imprecise sizer or imprecise adder).

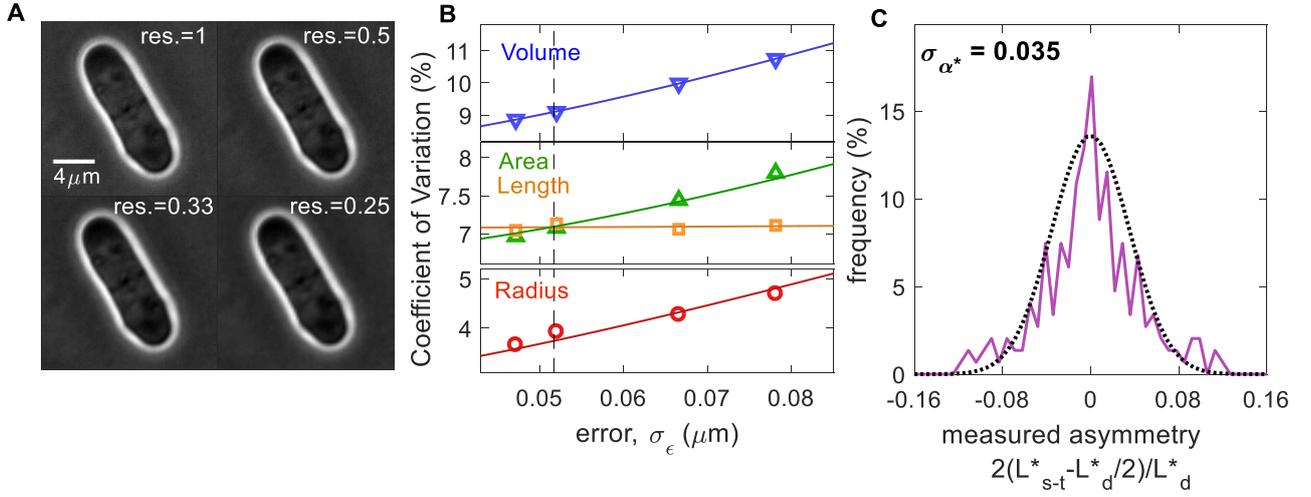

**Figure 2: Motivation and model parameter estimation.** (**A**) Reduction of image resolution by pixel binning: reported value of resolution is res=1/n, (where n x n pixels have been averaged and condensed into a single pixel; res=1 is the experimentally acquired image). (**B**) Fitting of the observed Coefficient of Variations at division (symbols) by tuning the parameters of the model (curves), at different image resolutions and therefore different experimental errors, $\sigma_\varepsilon$ (wild-type strain FC15, n=326). See also Fig. S1. Dashed line is located where $\sigma_\varepsilon = R\sigma_\rho$: for smaller errors, area sensing becomes apparent, i.e. $CV^2_{A^*_d} < CV^2_{L^*_d}$. (**C**) Measurement of division asymmetry and its fitting with a Gaussian distribution to determine $\sigma_\alpha$ (n=106), where $L^*_{s\text{-}t}$ is the measured distance between the division septum and cell tip. Experimental data are fitted with $\sigma_{\alpha*} = 0.035$. Since $\sigma^2_{\alpha*} \approx \sigma^2_\alpha + 5\sigma^2_\varepsilon/L^2$, we can extract $\sigma_\alpha = 0.032$ (Materials and Methods).

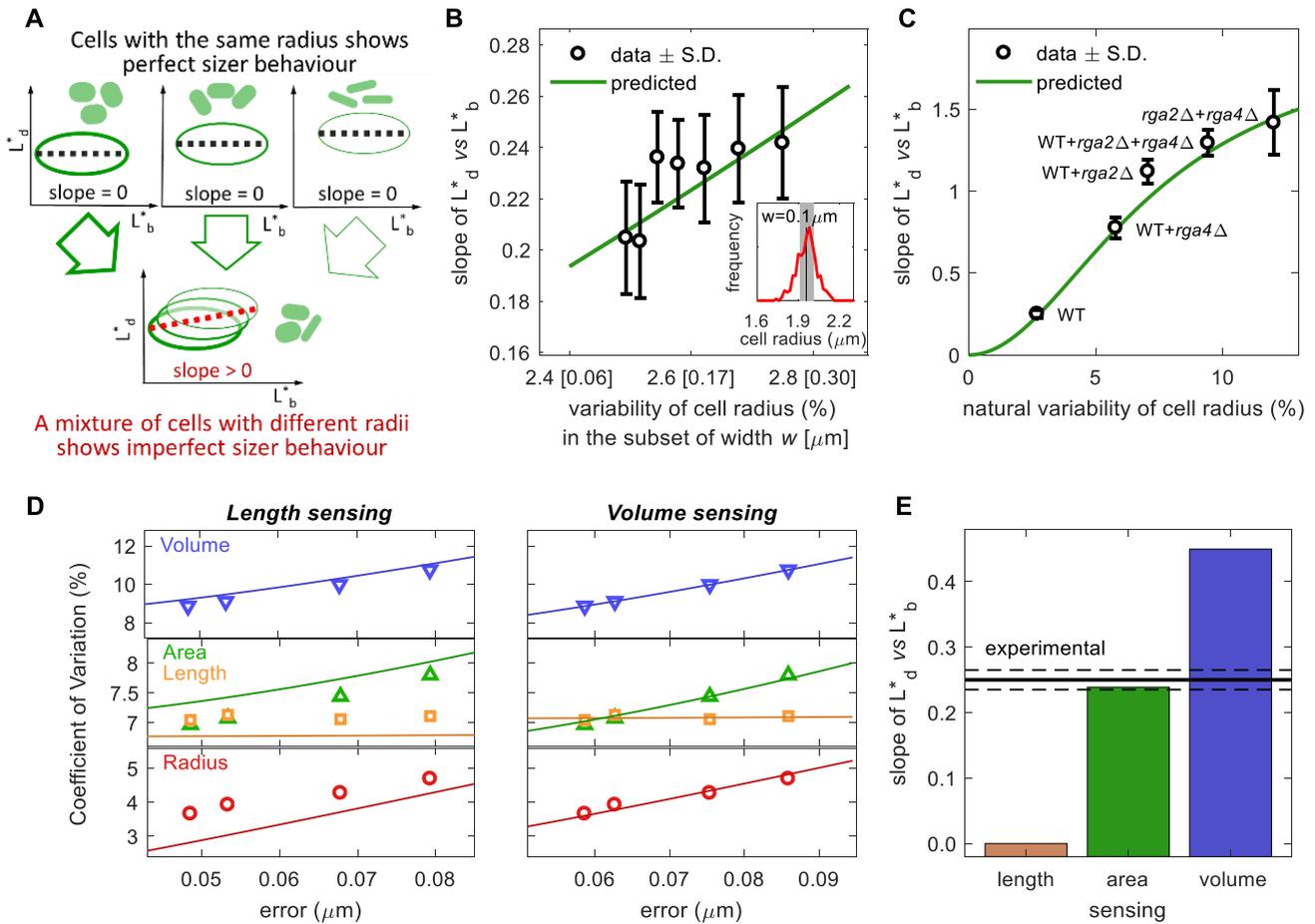

**Figure 3: Fission yeast implements a pure area-based sizer** (**A**) Intuitive explanation for how, for area sensing, variability in cell radius causes deviation from perfect sizer behaviour. Ellipses represent cloud of data points and dashed lines their linear fitting. (**B**) Inset: radius distribution in red and example of selection of cells with radius in range of ± $w$ = ± 0.1 μm around the mean value (grey shadow). Main plot: for given $w$ (value in squared brackets on x-axis), division-birth slope for length is measured on subset of selected cells (black circle, slope ± SD) and radius variability is calculated (Materials and Methods and Fig. S3B). Predicted slope based on area sensing is calculated with Eq. 2 (green curve). (**C**) Predicted area-based (green curve) and measured division-birth slope for length (mean value ± SD) obtained by combining mutants with different radii. (**D**) Fitting of the observed CV at division (symbols) by tuning the parameters of the model (curves), at different image resolutions and therefore different experimental errors, $\sigma_\varepsilon$. Left panels: length sensing model. Right panel: volume sensing model (see also Table S1). (**E**) Bars report the theoretical values of division-birth slope for length for each possible geometric size sensing model (based on length, area or volume). Area sensing is the one that provides the best agreement with the experimental data (black line ± SD, continuous and dashed lines). Panel D uses the same data as in Fig. 1A-C. For all other panels data for wildtype strain on agar (FC15, n=326) with automated segmentation.

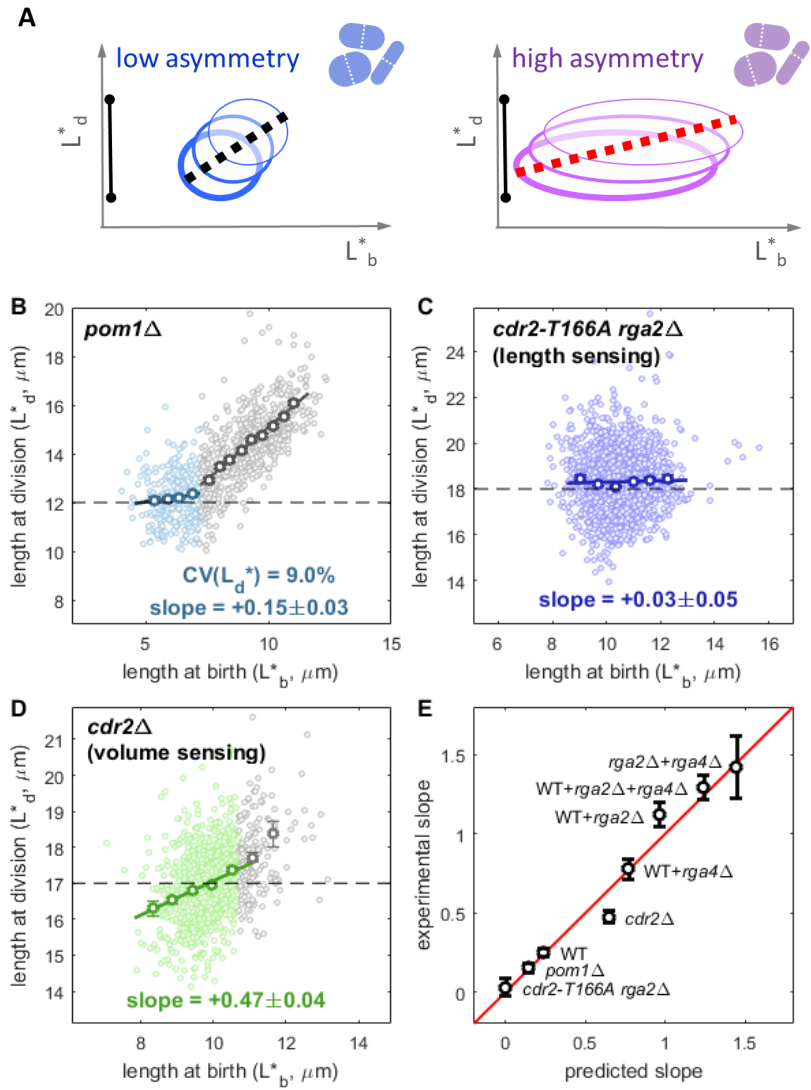

**Figure 4: Counter-intuitive predictions arising from asymmetric division errors and consequences of length or volume sensing.** (**A**) Possible explanation of how an increase in the asymmetry error at division brings the birth-division slope closer to zero. Black bars indicate the width of the distribution of the division length. (**B**) Size homeostasis plot for *pom1Δ* (FC2063, n=1757, data from [6]). Slopes of reported fit lines are 0.15±0.03 for the sizer regime (green colour, length at birth less than 7 μm) and +0.86±0.03 for the adder/timer-like regime (grey colour, length at birth greater than 7 μm). In the sizer regime, division length has a CV of 9.0%, higher than the wild-type, but has a lower birth-division slope (+0.15 versus +0.25 for the wild-type as reported in Fig. 1A), as predicted by the model. (**C**) Size homeostasis data for the *cdr2-T166A rga2Δ* mutant which implements length sensing shows near perfect sizer behaviour (FC3218, n=1785, data from [6]). (**D**) Size homeostasis data for the *cdr2Δ* mutant, which implements volume sensing. Regression line and reported slope of 0.47±0.04 refers to the binned data with length at birth smaller than 10 μm (green colour). Cells with length at birth greater than 10 μm (grey colour) show a slope of +0.88 ± 0.16 (FC3161, n=1046). Binned data (with mean value ± standard error) shown as bold circles in panels B-D. (**E**) Summary of the agreement between experimental slopes and values obtained by the model. Data points show the mean value ± standard error (see also Fig. S5B).

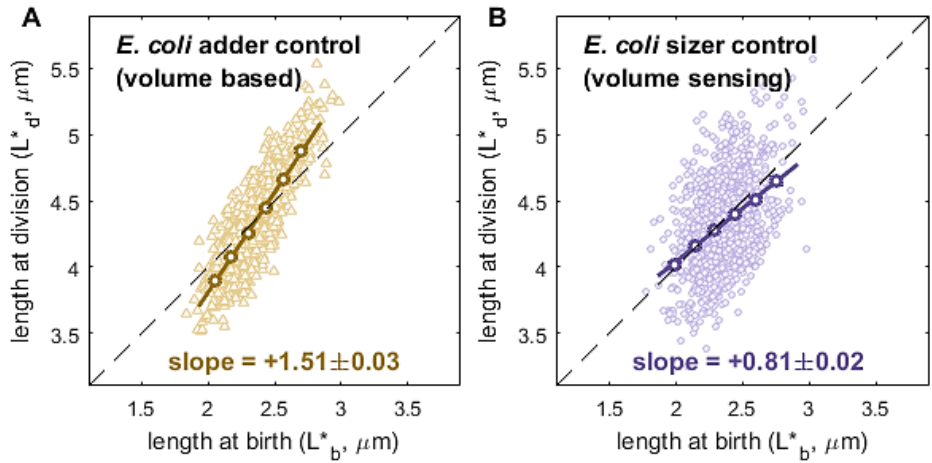

**Figure 5: Radius variability and sizer control reproduce the apparent adder behaviour in *E. coli*.** (**A**) Simulation of division-birth correlation based on length measurements for *E. coli* utilising pure adder control based on cell volume. (**B**) Same plot in case of pure sizer control based on cell volume. Both plots of panels A and B have been obtained with R=0.55 µm, V=3.77 µm³, 3.5% as the natural radius variability and the same noise used for fission yeast (Fig. S6B shows the effect of perturbations of these values). Binned data, with mean value ± standard error, also shown (dark circles), as well as best fit line, with indicated slope.

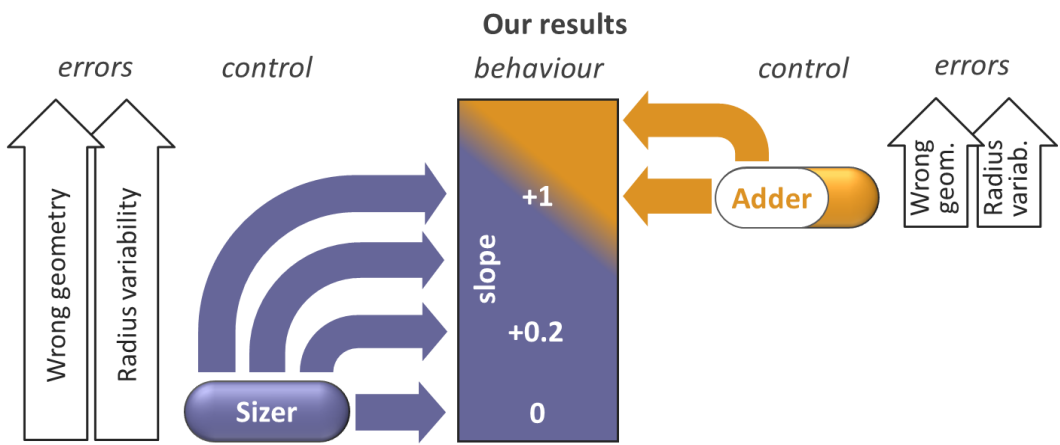

**Figure 6: Summary.** How a pure sizer mechanism can show a wide range of behaviours. Noise and error lead to an increase of the slope for sizer control (and also for adder control).

**TABLE 1: Variables and parameter values of the model.**

| Symbols | Description | units | Value for each strain ||||| 
|---|---|---|---|---|---|---|---|
| | | | WT | pom1Δ (subset of short cells only) | cdr2-T166A rga2Δ | cdr2Δ | E. coli |
| $\rho, \sigma_\rho$ | Gaussian random variable with standard deviation $\sigma_\rho$ (value reported here) describing cell-to-cell radius variability | % | 2.7 | 2.7 | 2.7 | 2.7 | 3.5 |
| $\mu_b, \mu_d, \sigma_\mu$ | Gaussian random variables with standard deviation $\sigma_\mu$ (value reported here) describing cell error in sensing its size | % | 6.5 | 8.6 | 6.5 | 6.5 | 6.5 [a] |
| $\alpha, \sigma_\alpha$ | Gaussian random variable with standard deviation $\sigma_\alpha$ (value reported here) describing the asymmetry in division, defined as $(2L^*_{s-t} - L^*_d)/L^*_d$ where $L^*_{s-t}$ is measured distance between the division septum and cell tip. | % | 3.2 | 3.2 | 3.2 | 3.2 | 3.2 [a] |
| $\varepsilon_b, \varepsilon_d, \sigma_\varepsilon$ | Gaussian random variables with standard deviation $\sigma_\varepsilon$ (value reported here) describing experimental error in the measurement at different resolutions res.=1 (0.0635 µm/pixel) res.=0.5 (0.127 µm/pixel) res.=0.33 (0.190 µm/pixel) res.=0.25 (0.254 µm/pixel) | µm | 0.045 0.052 0.067 0.078 | 0.06 [b] | 0.06 [b] | 0.06 [b] | 0.04 [b] |
| $R$ | Mean cell radius | µm | 1.96 | 1.89 | 1.58 | 1.95 | 0.55 |
| $L$ | Target cell length at division in case of length sensing (mean of the experimental measurement) | µm | n.a. | n.a. | 18.3 | n.a. | n.a. |
| $A$ | Target cell area at division in case of area sensing (mean of the experimental measurement) | µm² | 164 | 158 | n.a. | n.a. | n.a. |
| $V$ | Target cell volume at division in case of volume sensing (mean of the experimental measurement) | µm³ | n.a. | n.a. | n.a. | 183 | 3.77 |
| $q$ | Persistence of the cell radius from birth to division | - | 0.72 | 0.72 | 0.72 | 0.72 | - |

[a]: Simulations with additively increasing/decreasing each standard deviation by up to 1% (or 2%) are reported in Fig. S6.
[b]: Estimation based on the resolution of the images.

# Supplementary Information for

# Reassessment of the basis of cell size control based on analysis of cell-to-cell variability

Giuseppe Facchetti, Benjamin Knapp, Fred Chang, Martin Howard

## MATERIALS AND METHODS

**Image acquisition and analysis**

All images analysed here are taken from our previous publication [1]. High resolution images were used to obtain cell segmentation at different resolutions by binning pixels (Figs. 2, 3, S1). This allowed us to study the effect of the segmentation error. In these cases, we applied the following new semi-automated algorithm to the images acquired at the best z-plane. We first drew manually an approximate cell axis that connected the two cell tips (A and B, see Fig. S7A). The AB distance was a first estimation $L$ of the cell length. A first estimation of the cell radius was calculated as follows. From the middle point M of the AB segment, we derived an intensity profile along the direction orthogonal to the axis (towards both lateral borders of the cell, red dashed line in Fig. S7A). The steepest gradient of this signal identified the border of the cell, i.e. the cell radius. We repeated this gradient procedure using neighbouring points of M (at a distance k·dx, with k=-3, -2, ... , +2, +3, and where dx is the pixel size). The average of all these values gave our first estimation of the radius. We then shortened the segment AB by this amount from both ends. This defined the new segment A'B'. We divided this segment into n equal parts (n=0.5·L/dx) by placing n-1 internal points. The gradient procedure we used for the middle point M was then applied to all these points and to the two extremal points A' and B'. This identified the lateral borders of the cells. The symmetry axis of the resulting lateral borders was taken to be the new symmetry axis of the cell (segment CD). From both extreme points C and D, we derived the intensity profile along the different radial directions (from 0 to 180 degrees, steps of 2 degrees, purple dashed line in Fig. S7B). Ideally, the distance of the cell border to the CD segment should always have been equal to R for every point from C to D. However, cell irregularity and imaging defects altered this width. Therefore, after calculating the width from a point $x_i$, if the difference between this width and the width from the previous point $x_{i-1}$ was bigger than a given threshold (0.1 μm), we set the width equal to the width from $x_{i-1}$. The final result was a set of points that described the cell border. From these points we derived the cell length as the maximal distance between two points of the set (Fig. S7C). The cell radius was defined as the average of its value at different positions along the CD segment, i.e. excluding the two cell tips (Fig. S7C). Surface area and volume were then calculated using simple geometrical equations. As showed by Eq. 2 of the main text, segmentation error (i.e. high image resolution) is not critical in case of calculation of the division-birth slope based on length measurements (Fig. 1, 4, S2, S4B and S5). In these cases, phase contrast images were first analysed by a deep neural network machine learning algorithm [2]: this methodology generated binary images for feature (outline/cytoplasm) identification. These contours were then used for traditional gradient segmentation in Morphometrics [3]. This procedure is fully described in [1]. Strains used in the experiments are listed in Table S2.

**Equations for the pure sizer model**

Cells were assumed to have a perfect cylindrical shape with hemispherical ends. Surface area and volume were calculated using values of the cell length and of the cell radius: $A = 2\pi RL$ and $V = \pi R^2 L - 2\pi R^3/3$. We initially assumed no variation of the cell width during a single cell cycle, although later this assumption was relaxed.

**Notations:** In the following: $\mu$ is the relative error of the cell in sensing size, $\rho$ is the relative cell to cell variability of the radius (so that the true radius of a cell is $R(1+\rho)$), $\alpha$ is the relative error at division due to asymmetric misplacement of the septum (so that cell does not divide symmetrically in two halves but into $(1+\alpha)/2$ and $(1-\alpha)/2$ fractions), and $\varepsilon$ is the experimental error (in μm) in measuring distances. All these quantities are assumed to have a Gaussian distribution with mean zero. The following calculations will then be extended to negative values of the geometrical quantities but where the probability weight of these tails is too small to have a significant impact on the results. We also need different copies of some random variables in order to describe the variation at different moments of the cell cycle: for instance, depending on whether we are considering birth or division, the relative error of the cell in sensing size at division must be described by $\mu_b$ or $\mu_d$, respectively. These are independent and identically distributed random variables. The same holds for the error in our measurements ($\varepsilon_b$ for the length at birth, $\varepsilon_d$ for the length at division and $\varepsilon_r$ for the cell radius).

**Cell length, area and volume for the case of surface area sensing:** By stating that a cell is sensing surface area, this means that division occurs at a given area $A$ (plus the error $A\mu_d$ due to the cellular error in size sensing). Therefore, the real division length ($L_d$, not affected by our error in its measurement) of a cell with real radius $R(1+\rho)$ is the length such that the resulting cell area is equal to the target value corrected by the error due to the imperfect cell sensing, i.e. $2\pi R(1+\rho)L_d = A(1+\mu_d)$. By adding the error $\varepsilon_d$ of our measurement, we have the corresponding measured quantity $L_d^*$:

$$L_d^* = \frac{A(1+\mu_d)}{2\pi R(1+\rho)} + \varepsilon_d. \tag{S1}$$

Following similar reasoning, we derive the measured radius $R^* = R(1 + \rho) + \varepsilon_r$. Because of the experimental errors, the expression of the measured area at division does not coincide with the theoretical value corrected by the sensing error, i.e. $A(1 + \mu_d)$. By definition, the measured area at division is

$$A_d^* = 2\pi R^* L_d^* = 2\pi[R(1 + \rho) + \varepsilon_r]\left[\frac{A(1+\mu_d)}{2\pi R(1+\rho)} + \varepsilon_d\right], \tag{S2}$$

while the measured volume at division is

$$V_d^* = \pi(R^*)^2 L_d^* - \frac{2}{3}\pi(R^*)^3 = \pi[R(1 + \rho) + \varepsilon_r]^2 \left[\frac{A(1+\mu_d)}{2\pi R(1+\rho)} + \varepsilon_d\right] - \frac{2\pi[R(1+\rho)+ \varepsilon_r]^3}{3}. \tag{S3}$$

The length at birth is derived from the length at division (defined by replacing $\mu_d$ with $\mu_b$) and by including the error $\alpha$ due to asymmetric mispositioning of the division septum. After division, contraction of the ring and turgor pressure deform the plane of division into a new hemispherical end while conserving the radius and cell volume [4], leading to an extra term $R(1 + \rho)/3$:

$$L_b^* = \frac{A(1+\mu_b)(1+\alpha)}{4\pi R(1+\rho)} + \frac{R(1+\rho)}{3} + \varepsilon_b. \tag{S4}$$

We also verified that a use of a different fraction, e.g. $R(1 + \rho)/2$, does not affect our results. Similar to above, we also derived expressions for the measured area and the measured volume at birth. These equations allow us to calculate all the quantities measured in the experiments, namely the 4 coefficients of variation (for radius, length, area and volume), and the slopes of the 3 division-birth plots (using length, area or volume as the geometrical feature used for size control).

**Coefficients of variation:** The coefficient of variation of X is defined as $CV_X^2 = \sigma_X^2/\mathbb{E}[X]^2$, where $\mathbb{E}[X]$ denotes the expected value of the random variable *X*. As an example, we report here the calculation of the Coefficient of Variation (CV) of the measured length at division in the case of area size sensing, i.e. when:

$$L_d^* = \frac{A(1+\mu_d)}{2\pi R(1+\rho)} + \varepsilon_d.$$

Because of the small value of $\sigma_\rho$, the second-order approximation $(1 + \rho)^{-m} \approx 1 - m\rho + \frac{m(m+1)}{2}\rho^2$ is used in all calculations. The definition of the variance is

$$\sigma_{L_d^*}^2 = \mathbb{E}[(L_d^*)^2] - \mathbb{E}[L_d^*]^2.$$

Each term is then calculated as follows:

$$\mathbb{E}[(L_d^*)^2] = \iiint_{-\infty}^{+\infty} \left[\frac{A(1+\mu_d)}{2\pi R(1+\rho)} + \varepsilon_d\right]^2 P(\mu_d)P(\rho)P(\varepsilon_d)d\mu_d d\rho d\varepsilon_d =$$

$$\approx \left(\frac{A}{2\pi R}\right)^2 \iint_{-\infty}^{+\infty} (1 + 2\mu_d + \mu_d^2)(1 - 2\rho + 3\rho^2)P(\mu_d)P(\rho)d\mu_d d\rho + \int_{-\infty}^{+\infty} \varepsilon_d^2 P(\varepsilon_d)d\varepsilon_d$$

$$+ \iiint_{-\infty}^{+\infty} \frac{A(1+\mu_d)\varepsilon_d}{\pi R(1+\rho)} P(\mu_d)P(\rho)P(\varepsilon_d)d\mu_d d\rho d\varepsilon_d.$$

Since each Gaussian variable has zero mean value and retaining only lowest order terms, we have:

$$E[(L_d^*)^2] = \left(\frac{A}{2\pi R}\right)^2 (1 + \sigma_\mu^2 + 3\sigma_\rho^2) + \sigma_\varepsilon^2.$$

$$E[L_d^*] \approx \iiint_{-\infty}^{+\infty} \left[\frac{A}{2\pi R}(1 + \mu_d)(1 - \rho + \rho^2) + \varepsilon_d\right] P(\mu_d) P(\rho) P(\varepsilon_d) d\mu_d d\rho d\varepsilon_d = \frac{A}{2\pi R}(1 + \sigma_\rho^2).$$

$$\mathrm{CV}_{L_d^*}^2 = \frac{\sigma_{L_d^*}^2}{E[L_d^*]^2} = \frac{E[(L_d^*)^2] - (E[L_d^*])^2}{(E[L_d^*])^2} \approx \sigma_\mu^2 + \sigma_\rho^2 + \left(\frac{2\pi R}{A}\right)^2 \sigma_\varepsilon^2.$$

The same procedure was applied to all the other geometrical quantities (area, volume, radius). Here, we report only the final result for the other quantities. These expressions were then used to fit the data of Fig. 2B and extract parameter values for the errors.

$$\mathrm{CV}_{R^*}^2 = \sigma_\rho^2 + \frac{\sigma_\varepsilon^2}{R^2}.$$

$$\mathrm{CV}_{L_d^*}^2 \approx \sigma_\mu^2 + \sigma_\rho^2 + \left(\frac{2\pi R}{A}\right)^2 \sigma_\varepsilon^2.$$

$$\mathrm{CV}_{A_d^*}^2 \approx \sigma_\mu^2 + \left[\frac{1}{R^2} + \left(\frac{2\pi R}{A}\right)^2\right] \sigma_\varepsilon^2.$$

$$\mathrm{CV}_{V_d^*}^2 \approx \frac{A^2 R^2 \sigma_\mu^2 + (A^2 R^2 + 16\pi^2 R^6 - 8\pi A R^4)\sigma_\rho^2 + 4(A^2 + 5\pi^2 R^4 - 4\pi A R^2)\sigma_\varepsilon^2}{A^2 R^2 + \frac{16}{9}\pi^2 R^6 - \frac{8}{3}\pi A R^4}$$

We also investigated whether the square of the surface area CV is smaller than the square of the length CV:

$$\mathrm{CV}_{A_d^*}^2 - \mathrm{CV}_{L_d^*}^2 \approx \frac{\sigma_\varepsilon^2}{R^2} - \sigma_\rho^2.$$

In particular, we studied the sign of their difference. The right hand-side of the equation describes the linear relationship of the difference between the two CVs as a function of $\sigma_\varepsilon^2$. The negative intercept $-\sigma_\rho^2$ indicates that this difference can be negative if the error is sufficiently small. In particular, in order to have $\mathrm{CV}_{A_d^*}^2 - \mathrm{CV}_{L_d^*}^2 < 0$ we must have $\sigma_\varepsilon < R\sigma_\rho$, i.e. the error must not be bigger than the natural absolute variability of the radius (see Fig. 2B in the main text).

**Slopes:** We calculated the slopes for the plot of length at division ($L_d^*$) vs length at birth ($L_b^*$). The slope of the linear regression of a set of pairs $(x_i, y_i), i = 1, \ldots, N$ is

$$\mathrm{slope}(y, x) = \frac{N \sum x_i y_i - \sum x_i \sum y_i}{N \sum x_i^2 - (\sum x_i)^2},$$

which, for large $N$ and for our quantities ($L_d^*$ and $L_b^*$), can be rewritten as follows:

$$\mathrm{slope}(L_d^*, L_b^*) = \frac{E[L_b^* L_d^*] - E[L_b^*]E[L_d^*]}{E[(L_b^*)^2] - E[L_b^*]^2} = \frac{\mathrm{cov}(L_d^*, L_b^*)}{\mathrm{var}(L_b^*)}.$$

This expression explains why for the error in size sensing ($\sigma_\mu$), despite stretching along both the x- and y-axes, it is only the x-axis stretch that affects the division-birth slope. The above expectation values can be calculated using the same procedure adopted for the calculation of the CVs described above. With this analytical procedure, we derived Eq. 2 of the main text. As a verification of the approximations used, we also ran numerical simulations. In particular, we implemented a computational model which, by using Eqs. S1 and S4 and by simulating n=1000 cells, reproduced the same values of the analytical expressions (see Fig. S3A).

**Equations for models based on length sensing and on volume sensing for size control:** We report here the expressions of the initial quantities used to derive the CVs at division, and the division-birth slope, for the cases of length sensing and volume sensing, respectively (see above for the case of surface area sensing).

Length sensing:

$$L_d^* = L(1 + \mu_d) + \varepsilon_d.$$

$$A_d^* = 2\pi[R(1 + \rho) + \varepsilon_r][L(1 + \mu_d) + \varepsilon_d].$$

$$V_d^* = \pi[R(1 + \rho) + \varepsilon_r]^2[L(1 + \mu_d) + \varepsilon_d] - \frac{2\pi[R(1 + \rho) + \varepsilon_r]^3}{3}.$$

$$CV_{L_d^*}^2 \approx \sigma_\mu^2 + \frac{\sigma_\varepsilon^2}{L^2}.$$

$$CV_{A_d^*}^2 \approx \sigma_\mu^2 + \sigma_\rho^2 + \left(\frac{1}{R^2} + \frac{1}{L^2}\right)\sigma_\varepsilon^2.$$

$$CV_{V_d^*}^2 \approx \frac{9L^2\sigma_\mu^2 + 36(L - R)^2\sigma_\rho^2 + 9\left[1 + \frac{4}{R^2}(L - R)^2\right]\sigma_\varepsilon^2}{(3L - 2R)^2}$$

$$\text{slope}(L_d^*, L_b^*) = 0, \text{ always.}$$

Volume sensing:

$$L_d^* = \frac{V(1 + \mu_d)}{\pi R^2(1 + \rho)^2} + \frac{2R(1 + \rho)}{3} + \varepsilon_d.$$

$$A_d^* = 2\pi[R(1 + \rho) + \varepsilon_r]\left[\frac{V(1 + \mu_d)}{\pi R^2(1 + \rho)^2} + \frac{2R(1 + \rho)}{3} + \varepsilon_d\right].$$

$$V_d^* = \pi[R(1 + \rho) + \varepsilon_r]^2\left[\frac{V(1 + \mu_d)}{\pi R^2(1 + \rho)^2} + \frac{2R(1 + \rho)}{3} + \varepsilon_d\right] - \frac{2\pi[R(1 + \rho) + \varepsilon_r]^3}{3}.$$

$$CV_{L_d^*}^2 \approx \frac{9V^2\sigma_\mu^2 + 4(3V - \pi R^3)^2\sigma_\rho^2 + 9\pi^2 R^4\sigma_\varepsilon^2}{(3V + 2\pi R^3)^2}.$$

$$CV_{A_d^*}^2 \approx \frac{9V^2\sigma_\mu^2 + (3V - 4\pi R^3)^2\sigma_\rho^2 + \frac{1}{R^2}[(3V + 2\pi R^3)^2 + 9\pi^2 R^6]\sigma_\varepsilon^2}{(3V + 2\pi R^3)^2}.$$

$$CV_{V_d^*}^2 \approx \sigma_\mu^2 + \left[\frac{\pi^2 R^4}{V^2} + \left(\frac{2}{R} - \frac{2\pi R^2}{3V}\right)^2\right]\sigma_\varepsilon^2$$

$$\text{slope}(L_d^*, L_b^*) \approx \frac{2\left[\left(\frac{V}{\pi R^2}\right)^2 + \frac{2R^2}{9} - \frac{V}{\pi R}\right]\sigma_\rho^2}{\left(\frac{V}{\pi R^2} - \frac{2R}{3}\right)^2 \sigma_\rho^2 + \left(\frac{V}{2\pi R^2}\right)^2 \sigma_\mu^2 + \left(\frac{V}{2\pi R^2} + \frac{R}{3}\right)^2 \sigma_\alpha^2 + \sigma_\varepsilon^2}.$$

This last expression was used to calculate the slope for the *cdr2Δ* mutant (Fig. 4D) and for *E. coli*. In the former case, we used the same parameters values as for the wild-type, with a mean division length of 17 μm, whereas in the latter case we used parameters according to the data in the available literature [5]. In particular, we used the geometrical features of this bacterium ($R = 0.55$ μm, $V = 3.77$ μm³) and we estimated the natural variability $\sigma_\rho \approx 3.5\%$ from the value of the CV of the cell width in different growing media. To show the robustness of the result, we perturbed each noise up to ±2% and checked the distribution of the obtained slopes (Fig. S6B).

**Variability of the real radius in a subset of cells selected by the measured radius:** In Fig. 3B of the main text, we showed how the division-birth length slope reduces when the natural variability of the cell radius is reduced. To enact this strategy, we selected a subset of cells that have reduced variability. In particular we chose cells whose measured radius fell in the range $R \pm w$ (i.e. mean value ± w). In order to use Eq. 2 to calculate the predicted value of the division-birth slope for this subset of cells, we first needed to know the natural variability of this subset, which depends also on the experimental measurement error. We already know that

$$R^* = R(1 + \rho) + \varepsilon,$$

where R represents the average cell radius over the entire population. Suppose we have a cell with a given real radius $R_{\text{real}}$. First, we want to know the probability that the measured radius of this cell (i.e. $R_{\text{real}} + \varepsilon$) falls in $I_w = (R - w, R + w)$. This question is the equivalent of asking the probability that $\varepsilon$ belongs to the interval $(R - R_{\text{real}} - w, R - R_{\text{real}} + w)$. Because $\rho$ and $\varepsilon$ are independent random variables, we can just multiply this probability by the probability for a cell to have that real radius $R_{\text{real}}$, i.e. $\text{Prob}[R(1 + \rho) = R_{\text{real}}] = \text{Prob}[\rho R = R_{\text{real}} - R]$. Therefore, the probability that a cell of radius $R_{\text{real}}$ is selected:

$$p(R_{\text{real}}) = \frac{1}{N} \text{Prob}[\rho R = R_{\text{real}} - R] \text{Prob}[R - R_{\text{real}} - w < \varepsilon < R - R_{\text{real}} + w],$$

where $N$ is the normalization factor. This expression gives the distribution of the radius of the cell we select in $I_w$. The CV of this distribution then gives us the required natural variability.

For simplicity, we introduced $Z = R_{\text{real}} - R$ and rewrote the probability of a cell with radius $R_{\text{real}}$ to fall in $I_w$ as follows:

$$p(Z = z) = \frac{1}{N} \text{Prob}[\rho R = z] \text{Prob}[-w < \varepsilon + z < +w] = \frac{1}{N} \frac{1}{R\sigma_\rho \sqrt{2\pi}} e^{-\frac{z^2}{2R^2 \sigma_\rho^2}} \frac{1}{\sigma_\varepsilon \sqrt{2\pi}} \int_{-w}^{w} e^{-\frac{(q+z)^2}{2\sigma_\varepsilon^2}} dq$$

where the normalization factor is $N = \text{erf}\left(\frac{w}{\sqrt{2(R^2 \sigma_\rho^2 + \sigma_\varepsilon^2)}}\right)$.

By construction $\text{Var}[R_{\text{real}}] = \text{Var}[Z]$, which is equal to $\mathbb{E}[Z^2]$ because Z has zero mean value. This led to

$$\text{Var}[R_{\text{real}}] = R^2 \sigma_\rho^2 - \frac{2wR^4 \sigma_\rho^4}{R^2 \sigma_\rho^2 + \sigma_\varepsilon^2} \frac{e^{-\frac{w^2}{2(R^2 \sigma_\rho^2 + \sigma_\varepsilon^2)}}}{\sqrt{2\pi(R^2 \sigma_\rho^2 + \sigma_\varepsilon^2)}} \text{erf}\left(\frac{w}{\sqrt{2(R^2 \sigma_\rho^2 + \sigma_\varepsilon^2)}}\right)^{-1}.$$

The real natural variability is then the CV of $R_{\text{real}}$, i.e.

$$\text{Nat. var.}(w) = \sigma_\rho \sqrt{1 - \frac{2R^2\sigma_\rho^2}{R^2\sigma_\rho^2+\sigma_\varepsilon^2} \frac{w}{\sqrt{2\pi(R^2\sigma_\rho^2+\sigma_\varepsilon^2)}} \frac{e^{-\frac{w^2}{2(R^2\sigma_\rho^2+\sigma_\varepsilon^2)}}}{\text{erf}\left(\frac{w}{\sqrt{2(R^2\sigma_\rho^2+\sigma_\varepsilon^2)}}\right)}}. \tag{S5}$$

It is worth noticing that, because of the error, the accessible lower bound (in the limit $w \to 0$) of the natural variability is not zero, but rather $\text{Nat. var.}(w=0) = \sigma_\rho \sqrt{\frac{\sigma_\varepsilon^2}{R^2\sigma_\rho^2+\sigma_\varepsilon^2}}$. Equation S5 gives the curve of Fig. S3B and was used to calculate the natural variability of the experimental data in Fig. 3B of the main text.

**Equations for the pure adder model**

A similar approach was used also for adder control with incremental size $\Delta V$ defined by the volume. We report here only the equations for length at birth and length at division. Because several generations are needed to converge to the theoretical size, we implemented only the numerical version of the model. The simulation was run by starting with a length at birth equal to $\Delta V / \pi R^2$ and iterating 50 cell cycles. Analyses were performed on the cell size at the last cell cycle. Superscript [n] denotes a quantity at the *n*-th cell cycle.

$$L_b^{*[n]} = \frac{L_d^{[n-1]}(1+\alpha)}{2} + \frac{R(1+\rho)}{3} + \varepsilon_b.$$

$$L_d^{*[n]} = L_b^{[n]} + \frac{\Delta V(1+\mu_d)}{\pi R^2(1+\rho)^2} + \varepsilon_d.$$

From the numerical simulation of 3000 independent cells, we derived the value of the slope $L_d^*$ vs $L_b^*$.

# SUPPLEMENTARY FIGURES

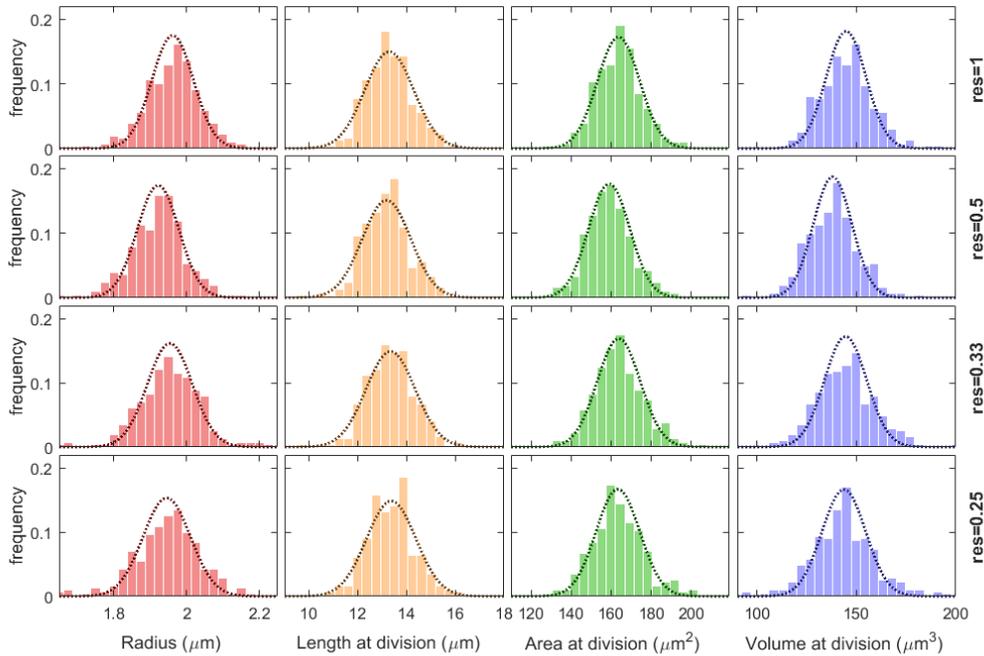

**Figure S1: Estimation of parameters.** Fitted distributions of the different geometrical quantities of cells at division for each resolution (Fig. 2A-B). Histogram bars: experimental data; dotted line: Gaussian distribution with variance obtained using the model equations.

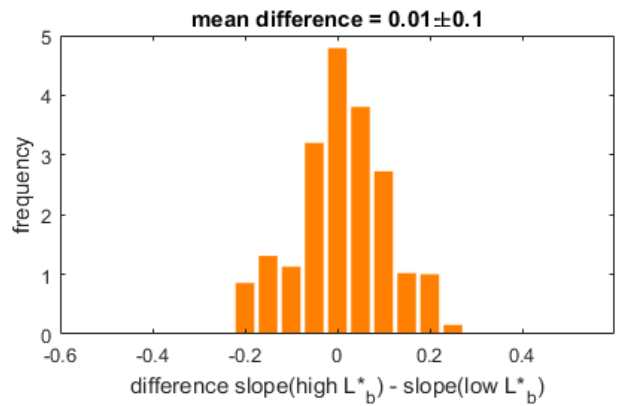

**Figure S2: Apparently imperfect sizer behaviour.** Slope of plot in Fig. 1A might be due to the presence of two regimes, i.e. a sizer regime (slope close to 0) at low birth lengths and an adder regime (slope close to 1) at high birth lengths (as in *pom1Δ*, see Fig. 4B in main text and [6]). To exclude this possibility, we split the cells into low and high birth lengths according to a threshold and calculated the slopes in the two regimes. We varied the value of the threshold and analysed the distribution of the difference between "slope at high $L_b$" and "slope at low $L_b$" (bar plot). Since the mean value of this distribution, as indicated in the figure, is very close to zero, we conclude that the cells do not show two-regime behaviour.

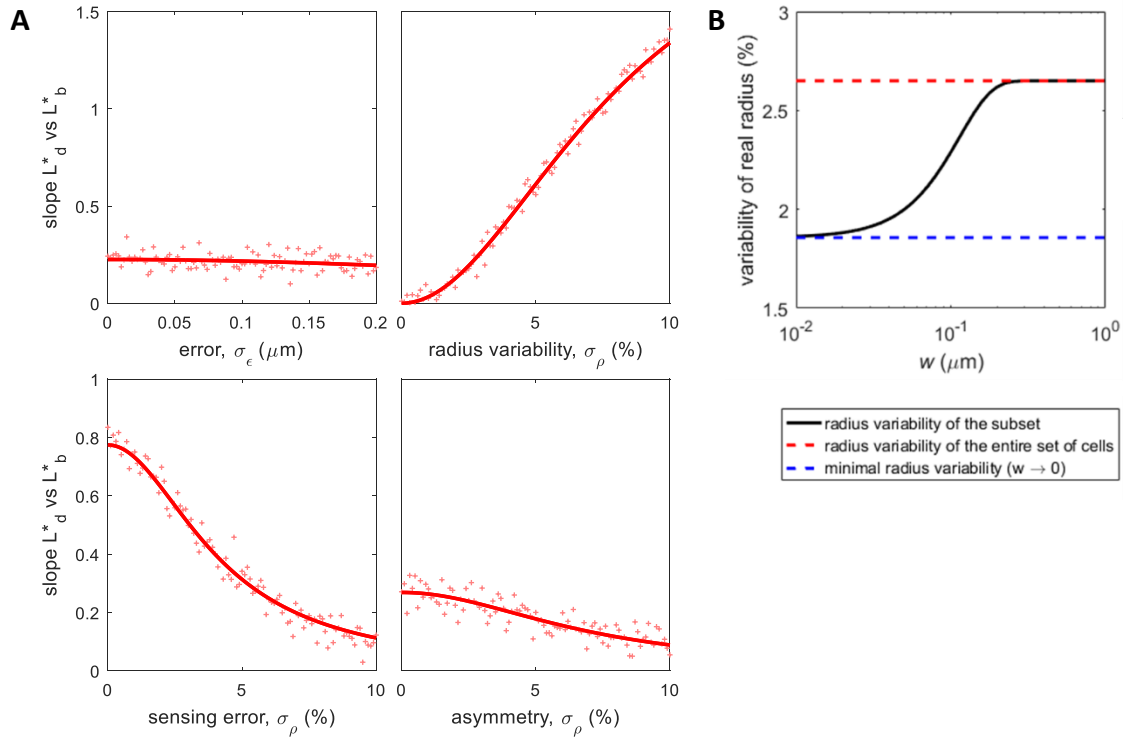

**Figure S3: Theoretical division-birth slopes for a pure area-based sizer control and effect of the noise sources.** (**A**) Effect of experimental error ($\sigma_\varepsilon$), radius variability ($\sigma_\rho$), sensing error ($\sigma_\mu$), asymmetry ($\sigma_\alpha$) on the division-birth slopes. Lines: analytical equations for division-birth slope ($L^*_d$ vs $L^*_b$), where subscript $d$ refers to division and subscript $b$ to birth, assuming underlying area-based size sensing. "+": numerical simulation (n=1000 cells). Except for the x-axis values, all other parameters are as in Table S1. (**B**) Variability of the real radius for the subset of cells with measured radius $R^*$ in the range $R \pm w$ (see Eq. S5). This analysis is used to derive the *x*-values in Fig. 3B.

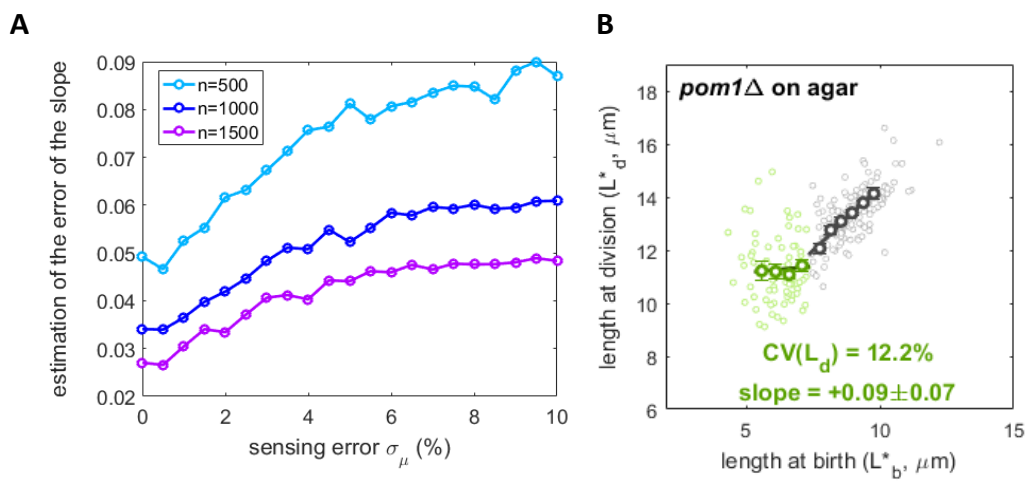

**Figure S4: Effect of sensing error $\sigma_\mu$ on division-birth slope uncertainty and repeated *pom1Δ* experiment.** (**A**) Estimation of the uncertainty of the division-birth slope as standard deviation over 1000 independent numerical simulations of n cells. To maintain the same error, a larger number of cells (n) must be imaged. (**B**) Size homeostasis plot for *pom1Δ*. Binned data, with mean value ± standard error, also shown (dark circles), together with best fit line. Slope of the second regime: +0.97±0.07. (FC2063, n=402).

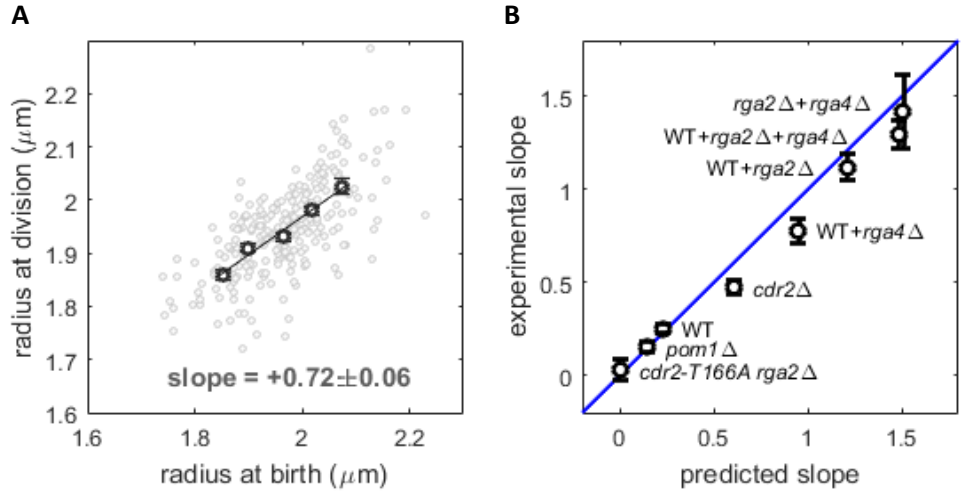

**Figure S5: Model with radius change during the cell cycle.** (**A**) Experimental data from wild-type cells (FC15, n=256). Binned data, with mean value ± standard error, also shown (dark circles), together with best fit line with slope $q$=0.72 and intercept 0.54, which is in good agreement with our theoretical value of $R(1-q) = 0.55$. (**B**) Comparison between experimental and predicted slope by using the model with radius changes. Data points show the mean value ± standard error. See also Fig. 4E.

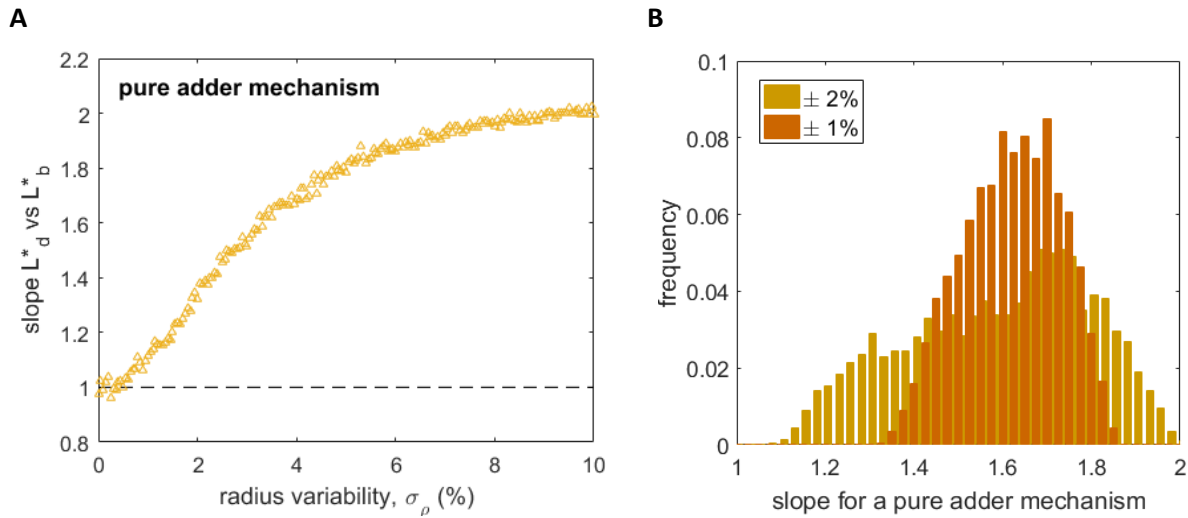

**Figure S6: Paradoxes in size behaviour.** (**A**) Effect of radius variability on the division-birth length slope in case of a pure adder mechanism based on volume control. Other parameters values: noise as for the other model predictions ($\sigma_\varepsilon = 0.04$ µm, $\sigma_\mu = 6.5\%$, $\sigma_\alpha = 3.2\%$), cell size $R = 0.55$ µm, $V = 3.77$ µm³ [5]. (**B**) Distribution of slopes when additively increasing/decreasing each standard deviation by up to 1% (or 2%) for all parameters of panel A, with fixed $\sigma_\rho = 3.5\%$.

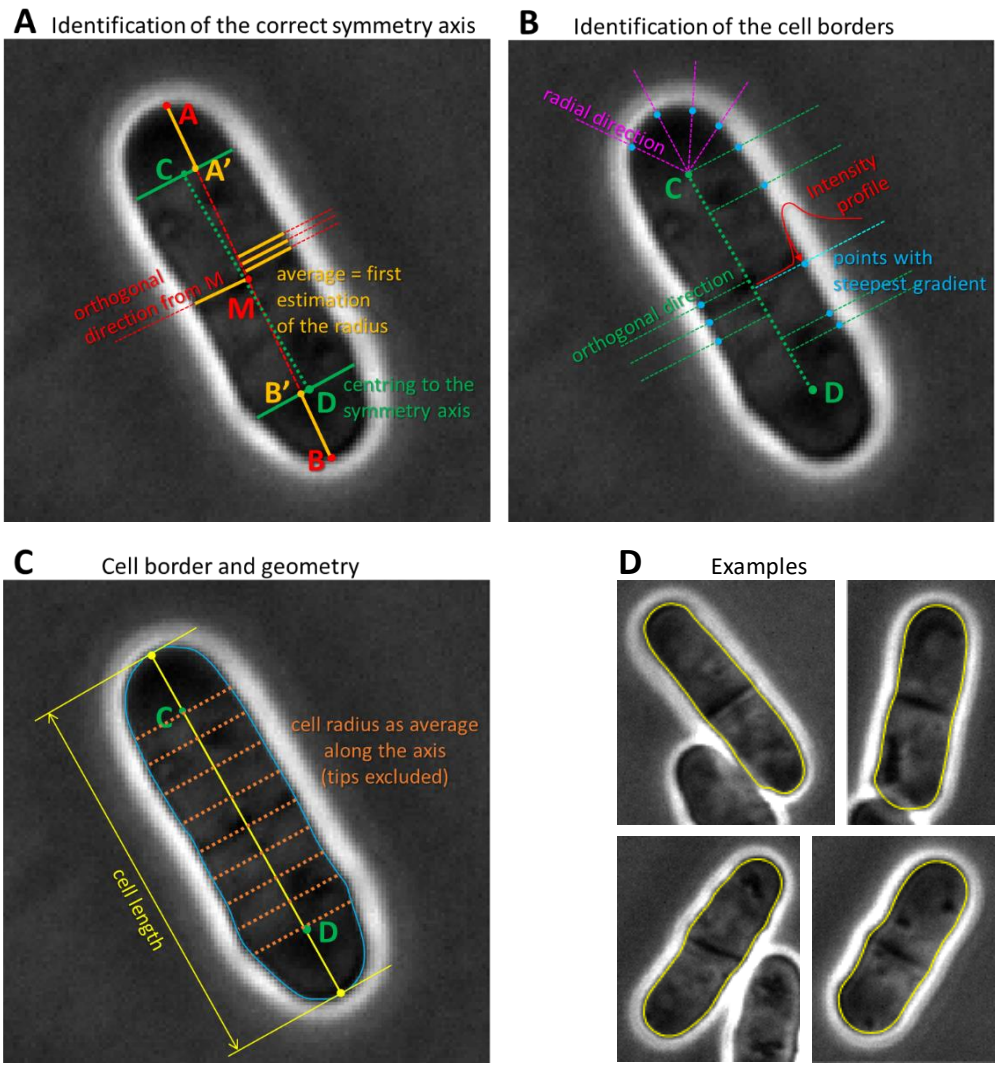

**Figure S7: Semi-automated segmentation algorithm**. (**A-C**) Sequence of operations performed to segment the cell border (see SI text). (**D**) Examples of segmented cells.

# SUPPLEMENTARY TABLES

**TABLE S1: Parameter values from fitting the CVs for each type of geometric size sensing** (Fig. 2B and Fig. 3D-E).

| parameter | $\sigma_\rho$ | $\sigma_\mu$ | $\sigma_\varepsilon$ (µm) at different resolutions | | |
|---|---|---|---|---|---|
| Length sensing | 1.7 % | 6.8 % | 0.036 | res.=1 | 0.0635 µm/pixel |
| | | | 0.043 | res.=0.5 | 0.127 µm/pixel |
| | | | 0.059 | res.=0.33 | 0.190 µm/pixel |
| | | | 0.072 | res.=0.25 | 0.254 µm/pixel |
| Area sensing | 2.7 % | 6.5 % | 0.045 | res.=1 | 0.0635 µm/pixel |
| | | | 0.052 | res.=0.5 | 0.127 µm/pixel |
| | | | 0.067 | res.=0.33 | 0.190 µm/pixel |
| | | | 0.078 | res.=0.25 | 0.254 µm/pixel |
| Volume sensing | 2.0 % | 6.8% | 0.058 | res.=1 | 0.0635 µm/pixel |
| | | | 0.062 | res.=0.5 | 0.127 µm/pixel |
| | | | 0.075 | res.=0.33 | 0.190 µm/pixel |
| | | | 0.085 | res.=0.25 | 0.254 µm/pixel |

**TABLE S2: Strains used in this study.**

| *S. pombe* strain | SOURCE |
|---|---|
| FC15: $h^-$ WT (972) | Lab collection |
| FC2947: $h^-$ rga2::ura4$^+$ ade6- leu1-32 ura4-D18 | Lab collection |
| FC1901: $h^-$ rga4::ura4$^+$ leu1-32 ura4-D18 | Lab collection |
| FC2063: $h^-$ pom1::natMX4 ade6- leu1-32 ura4-D18 | Lab collection |
| FC3161: $h^+$ cdr2::kanMX leu1-32 | Lab collection |
| FC3218: $h^-$ cdr2-T166A rga2::ura4$^+$ | Lab collection |

# SUPPLEMENTARY REFERENCES